\documentclass[pra, aps, twocolumn,floatfix,superscriptaddress,amsmath]{revtex4-1}
\usepackage{graphicx}
\usepackage{graphics}
\usepackage{braket}
\usepackage{mathtools}
\usepackage[titletoc]{appendix}
\usepackage{appendix}
\usepackage{float}
\def \bk{{\bf k}}
\def \br{{\bf r}}

\def \bj{{\bf j}}

\def \mum{\mu \mathrm{m}}
\def \mex{\mathrm{ex}}
\def \m2D{\mathrm{2D}}
\def \mB{\mathrm{B}}
\def \mms{\mathrm{ms}}
\def \mmms{\mathrm{mm/s}}
\def \mon{\mathrm{on}}
\def \moff{\mathrm{off}}

\def \ms{\mathrm{s}}

\def \cD{\mathcal{D}}

\def \cH{\mathcal{H}}

\def \matt{\mathrm{att}}
\def \mrep{\mathrm{rep}}

\def \msc{\mathrm{scale}}

\def \bk{\mathbf{k} }
\def \br{\mathbf{r} }

\begin{document}
\title{Sound propagation in a two-dimensional Bose gas across the superfluid transition}
\author{Vijay Pal Singh and Ludwig Mathey}
\affiliation{Zentrum f\"ur Optische Quantentechnologien and Institut f\"ur Laserphysik, Universit\"at Hamburg, 22761 Hamburg, Germany}
\affiliation{The Hamburg Centre for Ultrafast Imaging, Luruper Chaussee 149, Hamburg 22761, Germany}
\date{\today}
%
%
\begin{abstract}

Motivated by recent experiments in Phys. Rev. Lett. \textbf{121}, 145301 (2018), we study sound propagation in a two-dimensional (2D) Bose gas across the superfluid-thermal transition using classical field dynamics. 
Below the transition temperature we find a Bogoliubov and a non-Bogoliubov mode, above it we find the normal sound mode and the diffusive mode, as we determine from the dynamical structure factor. Our simulations of the experimental procedure agree with the measured velocities, and show that below the transition temperature the measurements detect the Bogoliubov mode. Above the transition, they either detect the normal sound mode for low densities or weak interactions, or the diffusive mode for high densities or strong interactions. As a key observation, we discuss the weak coupling regime in which the non-Bogoliubov mode has a higher velocity than the Bogoliubov mode, in contrast to a hydrodynamic scenario. We propose to detect this regime via step-pulse density perturbation, which simultaneously detects both sound modes. 

\end{abstract}

\maketitle
%
%

\section{Introduction}\label{sec:intro}

  Controlled excitation of quantum liquids has created insight into collective modes \cite{Jin1996, Ketterle1998, Foot2001, Chevy2002, Ferlanio2002, GrimmPRA2013, Salomon2015, Vale2017}, 
superfluidity \cite{Ketterle1999,Dalibard2000,Atherton2007,Anderson2010, Dalibard2012, 
Zwierlein2011, Miller2007, Weimer2015, Salomon2014, StratenPRL2009}, 
excitation properties \cite{Navan2019}, 
and sound diffusion \cite{Zwierlein2019}.
An intriguing phenomenon of quantum liquids is the propagation of second sound. 
It was first observed in liquid helium, where it is well described by Landau-Tisza's two-fluid hydrodynamic model \cite{Donnelly}. 
While the studies of superfluid helium were of crucial importance for understanding quantum liquids, 
the creation of Bose-Einstein condensates of dilute gases strongly expanded the scope of these studies. 
In ultracold quantum gases, a wide range of interactions and densities is available in a tunable manner, as well as bosonic and fermionic species, or mixtures thereof. 
The sound modes of 3D condensates were measured in Refs. \cite{Ketterle1997, Straten2009} and of a unitary Fermi gas  in Ref. \cite{Grimm2013}.
Theoretical studies were reported in Refs.  \cite{Griffin, Verney, Ilias}.

  Sound modes in 2D Bose gases are of special interest, as an interacting 2D system undergoes a superfluid transition via the Berezinskii-Kosterlitz-Thouless mechanism \cite{Minnhagen1987}. 
At the transition, and in the thermodynamic limit,  the superfluid density vanishes with a universal jump of $4/\lambda^2$, where $\lambda$ is the de Broglie wavelength. 
Furthermore, 2D systems exhibit a universal scale invariance: the dimensionless thermodynamic quantities, such as the phase-space density  and the entropy, depend only on a single dimensionless parameter $\mu/k_\mB T$ or equivalently $T/T_c$, where $\mu$ is the mean-field energy, $T$ the temperature, and $T_c$ the critical temperature.
This is confirmed by Refs. \cite{Chin, Dalibard}, 
where no jump in the thermodynamic quantities is observed.
Refs. \cite{Ozawa, Ota2018} studied the sound modes of 2D quasi-condensates using the two-fluid model,  
which show a jump at the transition.

     Recently, Ref. \cite{Dalibard2018} reported on the measurements of the sound propagation in a uniform 2D Bose gas of $^{87}$Rb atoms across the superfluid-thermal transition. 
The temperature dependence of the measured sound velocity shows no discernible jump in the crossover regime and a nonzero velocity above the transition. 
Theoretical studies of this measurement were reported in Refs. \cite{Ota, Salasnich}.

  In this paper, we investigate sound mode dynamics of a uniform 2D Bose gas of $^{87}$Rb atoms across the superfluid-thermal transition using c-field simulations.   
We determine a sound velocity $c$ by exciting running and standing waves with a weak Gaussian potential. 
These results show good agreement with the measurements of Ref. \cite{Dalibard2018}.
Below $T_c$, the temperature dependence of $c$ is captured by a Bogoliubov estimate that includes the superfluid density at nonzero temperature. 
Near and above $T_c$, $c$ displays a temperature dependence that depends on the density in a qualitative manner: $c$ increases and decreases for low and high density, respectively. 
This is also reflected in the dynamic structure factor, showing the density-dependent interplay between two sound modes that we refer to as the Bogoliubov and the non-Bogoliubov mode below $T_c$, and the diffusive and the normal sound mode above $T_c$. 
The results of $c$ show a breaking of the universal scale invariance at nonzero temperature due to Landau damping.
Going beyond the experimental work of Ref. \cite{Dalibard2018}, we propose to excite the two modes using a step-pulse density perturbation. 
The results of the step-pulse excitation across the transition show excellent agreement with the results of the dynamic structure factor, 
which provides a simultaneous measurement of both sound velocities.

\begin{figure}[]
\includegraphics[width=1.0\linewidth]{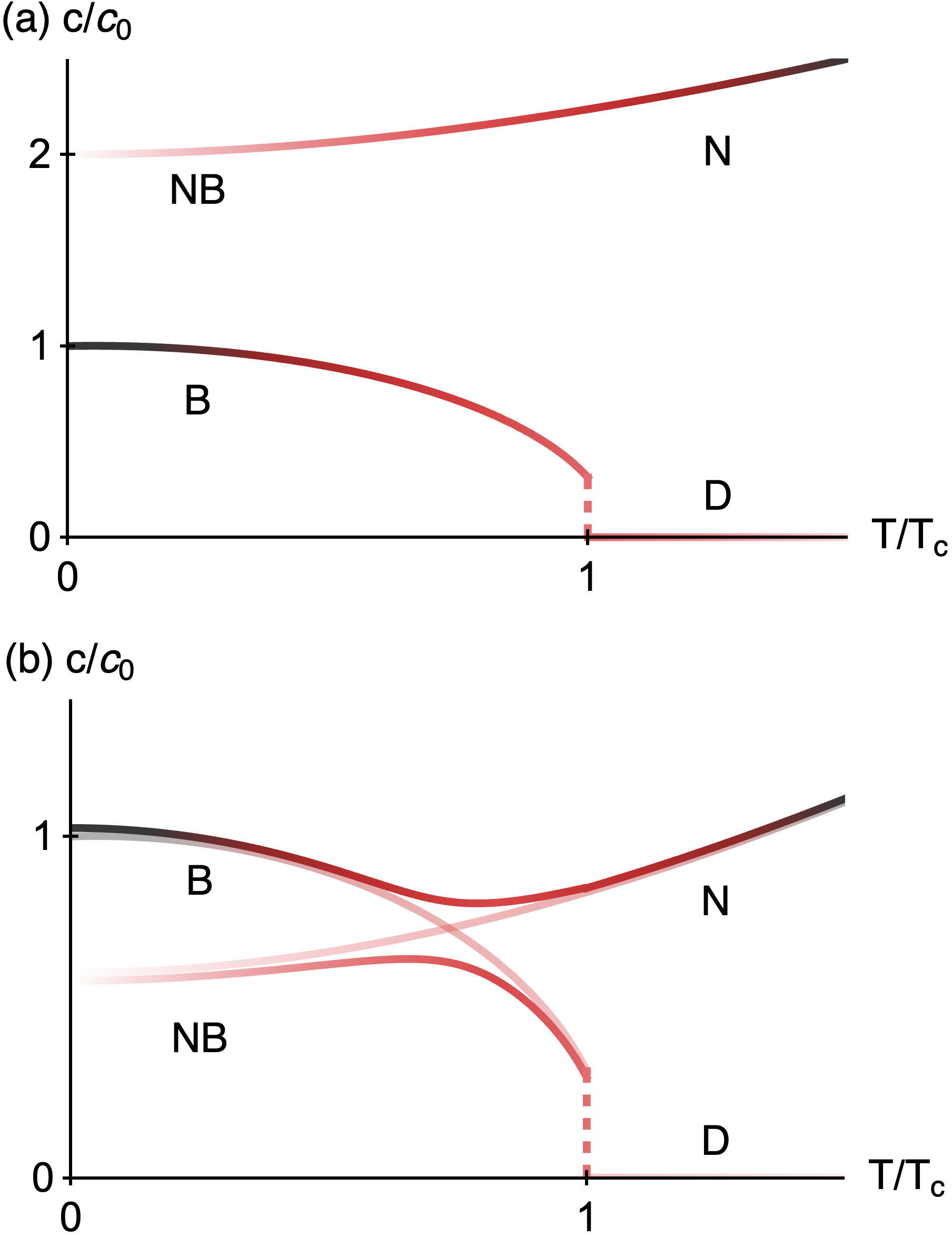}
\caption{Sketch of the qualitative temperature dependence of first sound (upper curve) and second sound (lower curve),  for (a) weak and (b) strong interactions. 
B labels the Bogoliubov sound mode, NB the non-Bogoliubov mode, D the diffusive mode, and N the normal sound mode. 
The line color represents the spectral weight of the modes in the dynamic structure factor.
For strong interactions the two modes undergo an avoided crossing at a hybridization temperature below the critical temperature, where the hybridization point is indicated by the crossing of the decoupled modes. }
\label{Fig:sketch}
\end{figure}

This paper is organized as follows. 
In Sec. \ref{sec:term} we illustrate the terminology of first and second sound. 
In Sec. \ref{sec_sim} we describe our simulation method.
In Sec. \ref{sec_density} we determine sound velocities by exciting running waves.
In Sec. \ref{sec_scale} we analyze the scale invariance of the sound velocity.
In Sec. \ref{sec_standing} we compare the running-wave velocity with the standing wave velocity and the Bogoliubov estimate. 
In Sec. \ref{sec_dsf} we show the dynamic structure factor.
In Sec. \ref{sec_two_sounds} we excite two sound modes with a step-pulse perturbation, 
and in  Sec. \ref{sec_conc} we conclude.

\section{Weak and strong coupling regime}\label{sec:term}

   In this paper, we refer to the faster mode as first sound, and the slower mode as second sound. We note that this terminology is inherited from the study of superfluid helium, and that the application of its terminology to cold atom systems could be done in several ways. 
As we describe in this paper,  for cold atom systems, we find two regimes. For weak interactions, or small densities, the temperature dependence of the sound velocities is sketched qualitatively in Fig. \ref{Fig:sketch}(a). At low temperatures, one sound mode is well described by the Bogoliubov approximation (B), and one mode that we refer to as a non-Bogoliubov (NB) mode. Here, the non-Bogoliubov mode is the first sound mode, in the sense that it is the faster mode. 
In Ref. \cite{Ilias} we have given a weak coupling description of this mode as a squeezing mode. 
For the interaction $g \rightarrow +0$, the ratio of the sound mode velocities approaches two. 
As the temperature is raised above the critical temperature, the NB mode continuously connects to the normal sound mode of a thermal gas. 
The velocity of the Bogoliubov mode undergoes a universal jump to zero at the critical temperature in the thermodynamic limit, and becomes the diffusive mode. 
We note that this sudden jump is replaced by a crossover regime for finite systems. 
The spectral weight of the modes in the dynamic structure factor has been indicated by the line color in the sketch.

  For strong interactions the temperature dependence of the mode velocities is sketched in Fig. \ref{Fig:sketch}(b). At low temperatures the Bogoliubov mode is the faster mode, which we refer to as first sound in this regime. 
The non-Bogoliubov mode is the slower mode, and connects to the diffusive mode. 
The two modes display an avoided crossing at a hybridization temperature below the critical temperature.
We give numerical evidence in support of these scenarios below.

\begin{figure*}[]
\includegraphics[width=1.0\linewidth]{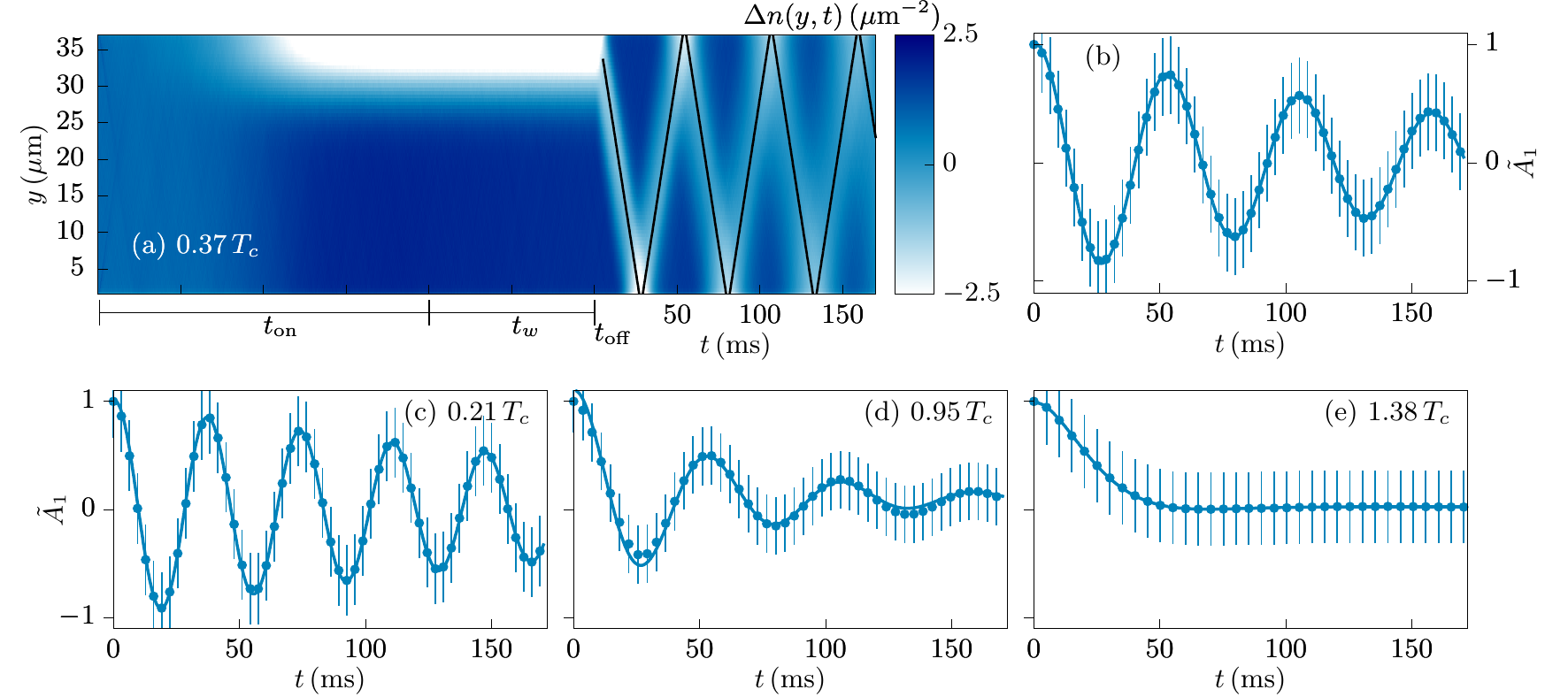}
\caption{\textbf{Running-wave excitation.} (a) Time evolution of the density profile $\Delta n(y, t)$ that is averaged along the $x$ direction, for $n_\m2D = 29.2 \, \mum^{-2}$ and $T/T_c=0.37$. 
We slowly turn on the Gaussian potential (smooth depletion of density along the upper edge), wait for $t_w$, and then turn it off at $t_\moff$, which excites a running wave, visible as a density dip propagating in space as a function of $t$.
The black continuous line is the triangular-wave function fit. 
Panel (b) shows the amplitudes $\tilde{A}_1(t)$ that are determined by fitting the density profile with the lowest-energy density mode. 
The errorbars are the standard deviation.
Panels (c)-(e) correspond to the parameter sets given in the text. 
The blue continuous lines are the fits to Eq. \ref{eq_damp}. }
\label{Fig1}
\end{figure*}

\section{Simulation method}\label{sec_sim}

 We simulate the dynamics of a 2D quasi-condensate using the c-field method of Ref. \cite{Singh2017}. 
We describe the system with the Hamiltonian
\begin{equation} \label{eq_hamil}
\hat{H}_{0} = \int d \br \, \Big[\frac{\hbar^2}{2m} \nabla \hat{\psi}^\dagger({\bf r})  \cdot \nabla \hat{\psi}({\bf r}) 
 + \frac{g}{2} \hat{\psi}^\dagger({\bf r})\hat{\psi}^\dagger({\bf r})\hat{\psi}({\bf r})\hat{\psi}({\bf r})\Big].
\end{equation}
$\hat{\psi}$ and $\hat{\psi}^\dagger$ are the bosonic annihilation and creation operator, respectively.
The 2D interaction parameter is given by $g = \tilde{g}\hbar^2/m$, where $\tilde{g}= \sqrt{8 \pi} a_s/\ell_z$ is the dimensionless interaction, $a_s$ the $s$-wave scattering length, and $\ell_z$ the harmonic oscillator length in the transverse direction. 
We use $\tilde{g} =  0.167$, as in Ref. \cite{Dalibard2018}.

  Inspired by the experimental setup of Ref. \cite{Dalibard2018}, 
we consider a 2D Bose cloud of $^{87}$Rb atoms confined in a rectangular box geometry of dimensions $L_x \times L_y = 34 \times 39 \, \mum $.  
For the numerical simulations, we discretize space with a lattice of size $N_x \times N_y = 68 \times 78$ and the discretization length $l= 0.5\, \mum$, where $l$ is chosen to be smaller than or comparable to the healing length $\xi$ and $\lambda$ (see Ref. \cite{Castin}).  
In our c-field approach, we replace the operators $\hat{\psi}$ in Eq. \ref{eq_hamil} and in the equations of motion by complex numbers $\psi$.
We sample the initial states in a grand-canonical ensemble of chemical potential $\mu$ and temperature $T$ via a classical Metropolis algorithm. We propagate this initial state according to the equations of motion. For each trajectory, we calculate the desired observables, and average over the initial thermal ensemble. The density of the atoms is in the range $n_\m2D = 3.0 - 53 \, \mum^{-2}$. 
For each $n_\m2D$, we choose several temperatures across the transition. 
The critical temperature $T_c$ is estimated by the critical phase-space density $\cD_c= \ln(380/\tilde{g})$ \cite{Prokofev2001}, 
which results in $T_c = 2\pi n_\m2D \hbar^2 /(m k_\mB \cD_c)$.

   To excite sound modes we add the perturbation $H_\mex= \int d \br V(\br, t) n(\br)$, 
where $n(\br)$ is the density at the location $\br = (x, y)$. The excitation potential $V(\br, t)$ is given by
\begin{align}\label{eq_pot}
V(\br, t) = V_0(t) \exp \bigl(- (y-y_0)^2/(2 \sigma^2)  \bigr),
\end{align}
where $V_0$ is the time-dependent strength and $\sigma$ the width. 
This potential is used along the upper edge of the box at the location $y_0= 36\, \mum$ to excite running and standing waves. For all simulations, $\sigma$ is $5\, \mum$ and $V_0$ is typically in the range $V_0/\mu = 0.1 - 0.4$, where $\mu = g n_\m2D$ is the mean-field energy. 
A running wave is excited using the following scheme. 
We slowly turn on the potential over $t_\mon= 200\, \mms$, i.e. $V_0(t_\mon)= V_0$, wait for $t_w = 100 \, \mms$, and then suddenly turn it off. 
This excites a sound wave propagating in space along $y$ direction as a function of $t$, see Fig. \ref{Fig1}(a). 
   To excite a standing wave, the following scheme is used. We slowly turn on the potential in the manner described above, and then sinusoidally modulate it by $V_0(t)= V_0(t^\prime)[1+ \sin\bigl(2\pi f(t-t^\prime) \bigr)]$, where $t^\prime = t_\mon + t_w$ and $f$ is the modulation frequency.
We perform this modulation at various frequencies $f$.  
After $1-2 \mathrm{s}$ excitation time, we analyze the density modulation following  Ref. \cite{Dalibard2018}. 
For each $f$, we explore one oscillation by recording the density profiles $n_i(y, f)$ at four different times $t_i$, $i \in \{1,2,3,4 \}$, 
where $t_i$ are chosen according to $\omega t_i = \omega t_1 + (i-1) \pi/2$, with $\omega = 2\pi f$. 
The amplitude of the standing waves is calculated by the quantity $q^2(y,f) = q_1^2(y,f) + q_2^2(y,f)$, where $q_1(y,f)= n_3(y, f) - n_1(y, f)$ and $q_2(y,f)= n_4(y, f) - n_2(y, f)$. 
The squared amplitudes determined at various $f$ are shown in Fig. \ref{Fig4}(a).   
We determine the sound velocities by exciting running and standing waves in Secs. \ref{sec_density} and \ref{sec_standing}, respectively.


\section{Sound propagation}\label{sec_density}

    In this section, we present the results of running-wave excitation for various combinations of $n_\m2D$ and $T/T_c$.
As an example, we first choose $n_\m2D = 29.2 \, \mum^{-2}$ and $T/T_c=0.37$, which is one parameter set used in the experiment. 
We excite a running wave following the sequence described in Sec. \ref{sec_sim}.  
In Fig. \ref{Fig1}(a) we show the time evolution of the density profile $\Delta n(y, t)= n(y,t)- n_\m2D$, 
which is averaged over the $x$ direction and the ensemble.
The excited sound wave is indicated by the density dip propagating in space as a function of time.
The sound wave travels back and forth between the edges at a constant velocity and forms a triangular pattern.
We fit the locations of the sound wave with a triangular-wave function to determine its velocity $c$. 
From the fit, we obtain $c=1.47 \, \mmms$, which is in excellent agreement with the measured $c = 1.49 \, \mmms$. 
The simulated $c$ is slightly below the Bogoliubov estimate of the sound velocity at zero temperature 
$c_0 =\sqrt{g n_\m2D /m}= 1.61\, \mmms$.

   Furthermore, we examine the damping of the sound mode in Fig. \ref{Fig1}(a). 
We fit the density profile with the function $n(y,t) = \bar{n} + A_1(t) \cos(\pi y/L_y)$ to determine the amplitudes $A_1$. 
This function represents the lowest-energy density mode, and the functional form is motivated by the experiments. 
$\bar{n}$ is a fitting parameter and represents the average density in $y$ direction.
In Fig. \ref{Fig1}(b) we show the extracted amplitudes $\tilde{A}_1 (t) = A_1(t)/A_1(0)$ as a function of the propagation time $t$. 
We fit $\tilde{A}_1$ with an exponentially damped sinusoidal function \cite{Dalibard2018}
\begin{align}\label{eq_damp}
f(t) =  e^{-\Gamma t/2} \bigl[\Gamma/(2\omega) \sin(\omega t) +   \cos(\omega t)  \bigr]
\end{align}
to determine the frequency $\omega$ and the damping rate $\Gamma$. 
From these, we determine the sound velocity $c= \omega L_y/\pi $ and the quality factor $Q= 2 \omega/ \Gamma$. 
For $\tilde{A}_1$ in Fig. \ref{Fig1}(b), the fit yields $\omega = 119.7\, \ms^{-1}$ and $\Gamma = 11.2\, \ms^{-1}$, 
and we obtain $c= 1.47 \, \mmms$ and $Q= 21.4$. 
The value of $c$ is the same as for the triangular pattern fit, and the high value of $Q$ implies weak damping of the sound mode. As the main origin of the damping of the sound modes, we identify Landau damping as we explain below.


\begin{table}[ ]
\begin{tabular}{|c| c| c| c| c| c| c| c| c|}
\hline 
\multicolumn{3}{|c|}{Parameter set} & \multicolumn{3}{|c|}{Simulation} &  \multicolumn{3}{|c|}{Experiment}\\
\cline{1-9}  
$n_\m2D $ &  $T/T_c$    & $V_0/\mu$  & $c\, [\mmms]$  & $\Gamma  $  &
$Q$  &  $c\, [\mmms]$ & $\Gamma $ & $Q$ \\[0.5ex] 
\hline 
$53\, \mum^{-2}$ & 0.21 & 0.2 & 2.10 & 8.7  & 39.1 & 2.20 & 5.5 & 64.5  \\ 
$52\, \mum^{-2}$ & 0.95 & 0.2 & 1.46 & 35.8 & 6.6 &  1.53 & 31.0 & 11.1  \\
$11\, \mum^{-2}$ & 1.38 & 0.4 & 0.56 & 111  & 0.8 &  0.81 & 32.7 & 4.2 \\ [1ex] 
\hline 
\end{tabular}
\caption{Values of $c$,  $\Gamma$, and $Q$, obtained from the simulation,  compared to the measurements in Ref. \cite{Dalibard2018}, for the same parameter sets of  $n_\m2D$ and $T/T_c$. 
$V_0$ is the strength of the Gaussian potential. 
$\Gamma$ is given in units of $\ms^{-1}$. }
\label{tab_c3} 
\end{table}

 \begin{figure}[]
\includegraphics[width=1.0\linewidth]{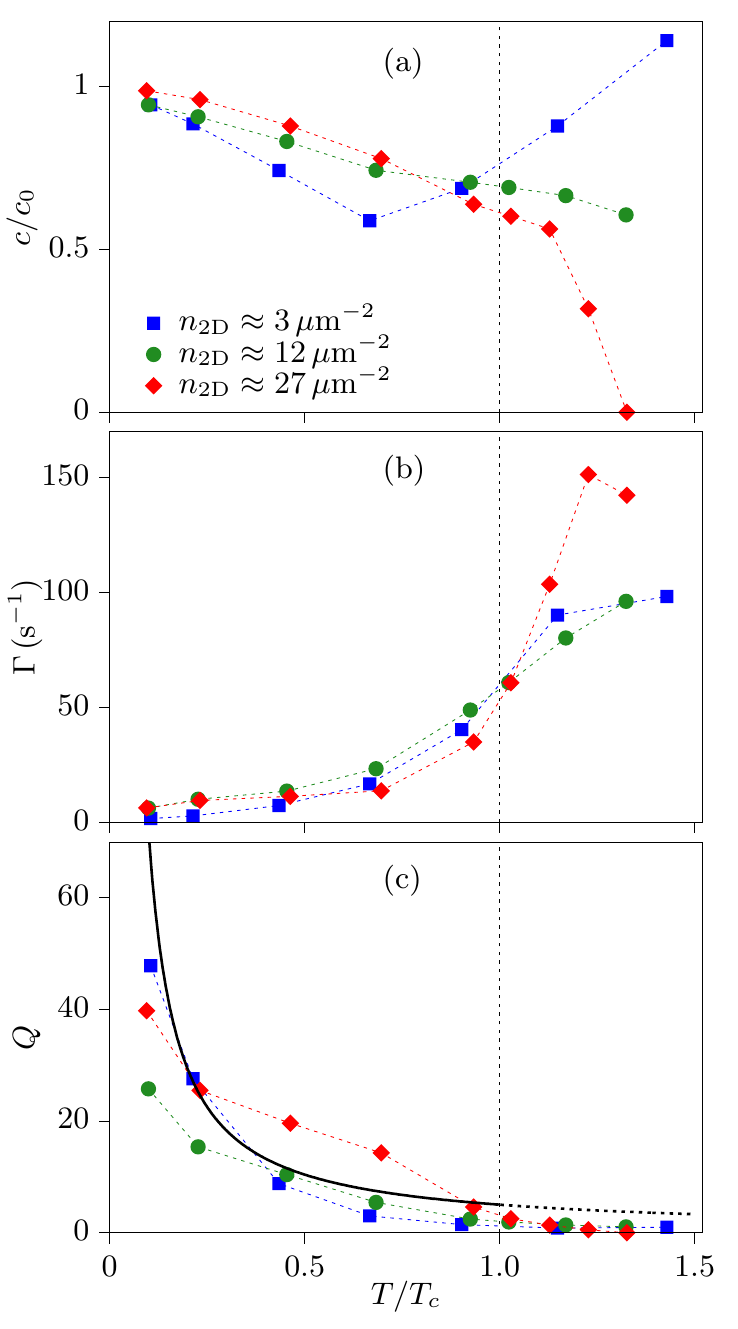}
\caption{\textbf{Sound velocity, damping rate, and quality factor. }  
(a) Normalized sound velocity $c/c_0$ as a function of $T/T_c$ for low (blue squares), moderate (green circles), and high $n_\m2D$ (red diamonds).
Damping rate $\Gamma$ and quality factor $Q$ are shown in panels (b) and (c), respectively. 
The black continuous line in panel (c) is the Landau prediction \cite{Dalibard2018}. 
The vertical dashed line at $T/T_c = 1$ denotes the critical point \cite{Prokofev2001}. 
The thin dashed lines are guides to the eye. }
\label{Fig2}
\end{figure}

   We now consider the three other sets of $n_\m2D$ and $T$ that are used in the experiment, which are $(53 \, \mum^{-2}, 0.21\, T_c)$, $(52 \, \mum^{-2}, 0.95\, T_c)$, and $(11 \, \mum^{-2}, 1.38\, T_c)$. 
For each set, we repeat the running-wave excitation and determine $\omega$ and $\Gamma$, as above. 
We show the extracted amplitudes $\tilde{A}_1$ in Figs. \ref{Fig1}(c), \ref{Fig1}(d), and \ref{Fig1}(e), respectively. 
The values of $c$, $\Gamma$, and $Q$ are given in Table \ref{tab_c3}, 
where we compare them with their corresponding measured values. 
They are in agreement below $T_c$, while they  deviate for the parameter set above $T_c$. 
We link this deviation to the measurement uncertainty and possibly different values of $V_0$ between experiment and simulation.

  We now analyze the temperature dependence of $c$ across the transition systematically. 
We choose the three densities $n_\m2D \approx 3$, $12$, and $27 \, \mum^{-2}$.
We refer to them as low, moderate, and high density, respectively.
For each $n_\m2D$, we determine $c$, $\Gamma$, and $Q$ at various $T/T_c$, with the running-wave excitation described above. 
We use the same $V_0 \approx 0.2\, \mu$ for all simulations.
We show the normalized results of $c/c_0$ as a function of $T/T_c$ in Fig. \ref{Fig2}(a). 
The temperature range includes the superfluid, crossover, and thermal regime. 
In the superfluid regime, $c$ overall decreases with increasing $T$. 
The reduction in $c/c_0$ is higher for low $n_\m2D$ as compared to high $n_\m2D$. 
In the crossover and the thermal regime the temperature dependence of $c/c_0$ depends on the density in a qualitative manner. 
With increasing $T/T_c$, $c/c_0$ increases for a small density $n_\m2D$, but decreases for large density. 
Note that $c$ eventually vanishes in the thermal regime for high $n_\m2D$. 
This result indicates that at all densities, the running-wave measurement primarily excites the Bogoliubov mode at temperatures below the transition temperature. However, above the transition temperature, the potential quench primarily excites the normal sound mode at low density, or weak-coupling, and the diffusive mode at high densities. 
As we describe below, the same trend is visible in the standing-wave experiment. 
Furthermore, the dynamic structure factor that we discuss in Sec. \ref{sec_dsf} supports this scenario as well. 
We emphasize that, in general, both modes are excited in these experiments. However, the amplitudes of the excited states are in general very different so that only one mode is detectable. In Sec. \ref{sec_two_sounds} we present a proposal for exciting both modes simultaneously with detectable amplitudes.

  In Figs. \ref{Fig2}(b) and \ref{Fig2}(c) we show the damping rate $\Gamma$ and the quality factor $Q$, respectively. 
$\Gamma$ shows a density-dependent behavior as a function of $T/T_c$, which translates into a density dependence of the temperature 
dependence of $Q$. As a comparison we depict the prediction for the $Q$ factor, 
which assumes that Landau damping is the primary mechanism for the line broadening, see Refs. \cite{Dalibard2018, Chung}.
The comparison shows good agreement.

\begin{figure}[]
\includegraphics[width=1.0\linewidth]{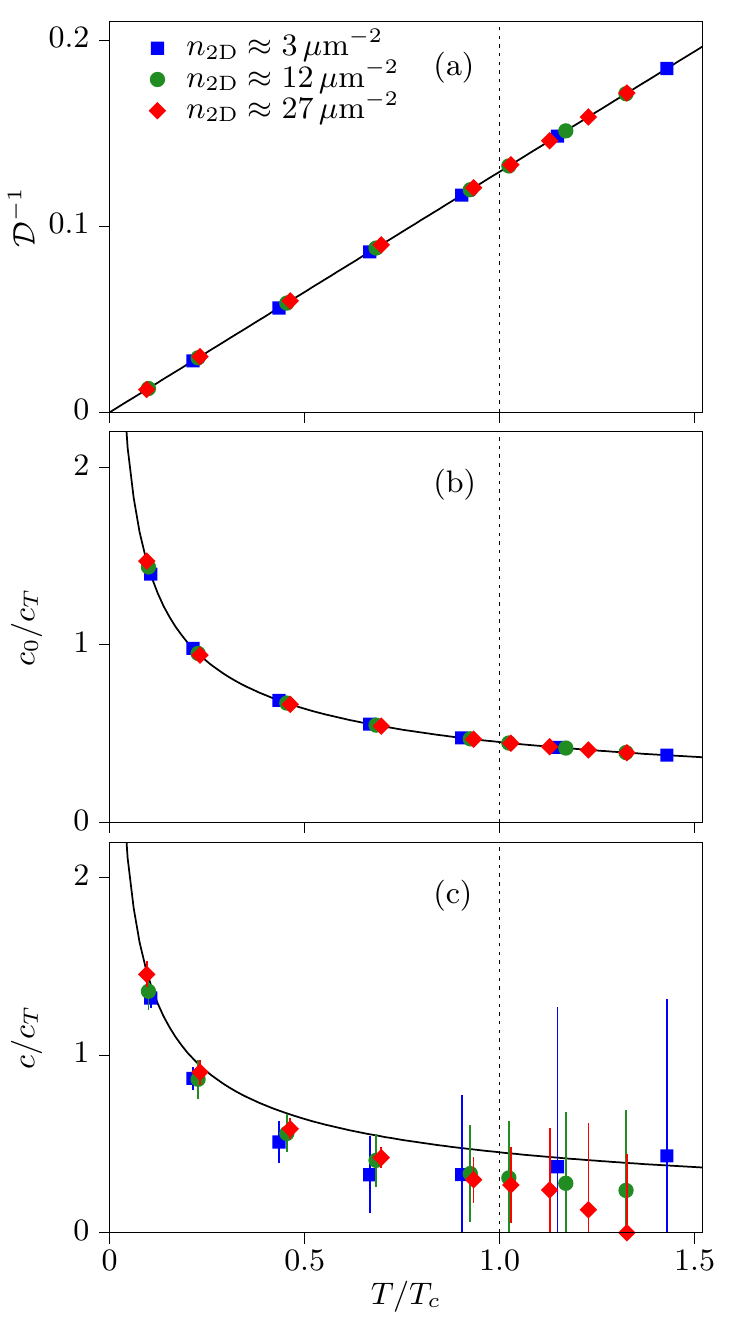}
\caption{\textbf{Scale invariance. }  (a) Inverse phase-space density $\cD^{-1}$ plotted against $T/T_c$ for low  (blue squares), moderate (green circles), and high $n_\m2D$ (red diamonds). 
The black continuous line is the scaling prediction $\cD^{-1}_\msc$. 
Panels (b) and (c) show the results of $c_0/c_T$ and $c/c_T$, respectively. 
The  black continuous lines in panels (b) and (c) are the scaling prediction $(c_0 /c_T)_\msc$. 
The errorbars in panel (c) represent the damping velocities $\Gamma L_y/\pi$ determined with $\Gamma$ in Fig. \ref{Fig2}(b).  }
\label{Fig21}
\end{figure}

\section{Scale invariance} \label{sec_scale}
        
  Here we examine the scale invariance of $c$ across the transition. 
We first demonstrate the scale invariance of the phase-space density $\cD= n_\m2D \lambda^2$. 
We calculate $\cD$ at various $T/T_c$ for the same three densities as before.   
In Fig. \ref{Fig21}(a) we show  the results of the inverse phase-space density $\cD^{-1}$ determined at various $T/T_c$.  
The different $n_\m2D$ results collapse on a single line all across the transition.
We compare them with the scaling prediction $\cD^{-1}_{\msc} = \cD_c^{-1}T/T_c$, where $\cD_c$ is the critical phase-space density \cite{Prokofev2001}. 
The simulations are in excellent agreement with the prediction. 
This confirms the universal scale invariance of the phase-space density.

    We now test the scale invariance of the dimensionless sound velocities $c_0/c_T$ and $c/c_T$, 
where we refer to $c_T = \sqrt{k_\mB T/m}$ as the thermal velocity.
The scaling prediction for $c_0/c_T$ is $ (c_0/c_T)_\msc = \sqrt{ \cD_c \tilde{g} T_c/(2\pi T)}$, 
which depends only on $T/T_c$, while $\tilde{g}$ is a fixed parameter. 
We show the results of $c_0/c_T$ and $c/c_T$ in Figs. \ref{Fig21}(b) and \ref{Fig21}(c), respectively.  
The results of $c_0/c_T$ collapse on a single line and agree very well with $(c_0/c_T)_\msc$.
This is a direct consequence of the data collapse shown in Fig. \ref{Fig21}(a). 
However, for  $c$,  obtained from the simulation, the different $n_\m2D$ results do not collapse on a single line, 
which shows a breaking of scale invariance regarding the sound velocity. 
The results of $c/c_T$ and the prediction $(c_0/c_T)_\msc$  agree only at low $T$, 
whereas they deviate at intermediate and high $T$. 
At low $T$, the damping of the sound mode is small compared to the mode frequency, i.e. $\Gamma \ll \omega$. 
However, at high $T$, $\Gamma$ is comparable to $\omega$ and the deviation from the scaling prediction increases. 
The magnitude of the damping is expressed as a velocity, and shown as errorbars in Fig. \ref{Fig21}(c). 
The deviation from the scaling prediction is comparable to the errorbars, suggesting that this breaking of scale invariance is due to the damping of the sound mode.
Near and above $T_c$, $c$ undergoes the density-dependent changes that we have pointed out in the previous section.

\section{Standing waves}\label{sec_standing}

\begin{figure}[]
\includegraphics[width=0.95\linewidth]{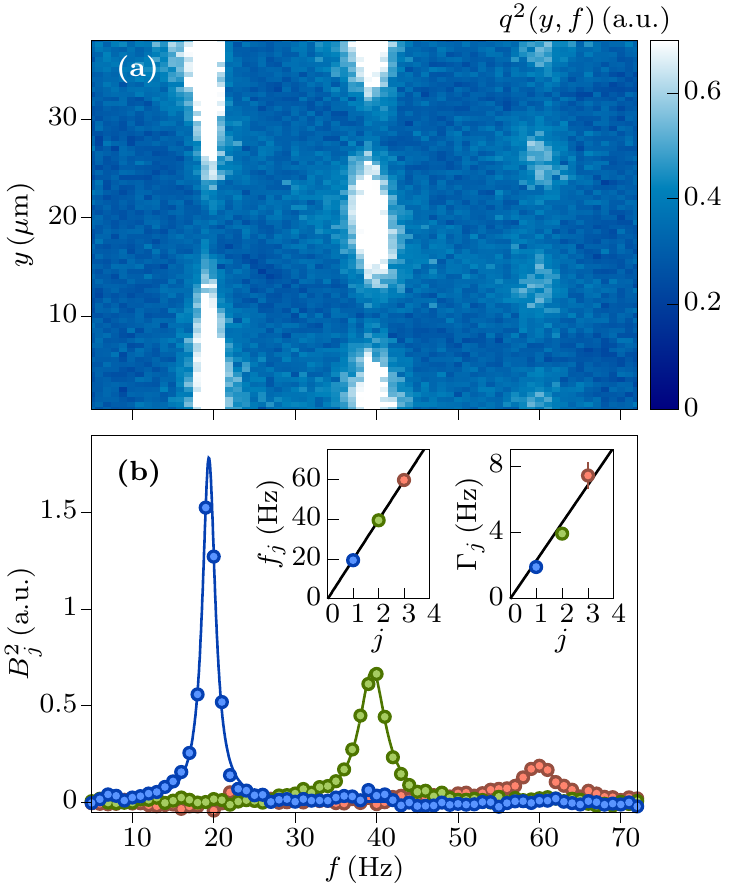}
\caption{\textbf{Standing-wave excitation. } 
(a) Squared amplitude $q^2(y, f)$ calculated at varying modulation frequency $f$, 
for $n_\m2D \approx 27 \, \mum^{-2}$ and $T/T_c=0.23$. 
Panel (b) shows the determined amplitudes $B_j^2$ of the standing waves, where $j$ is the mode index. 
The continuous lines are the Lorentzian fits. 
The mode frequencies $f_j$ and the damping rates $\Gamma_j$ are given in the insets, 
where the continuous lines are the linear fits.   }
\label{Fig4}
\end{figure}

\begin{figure}[t]
\includegraphics[width=1.0\linewidth]{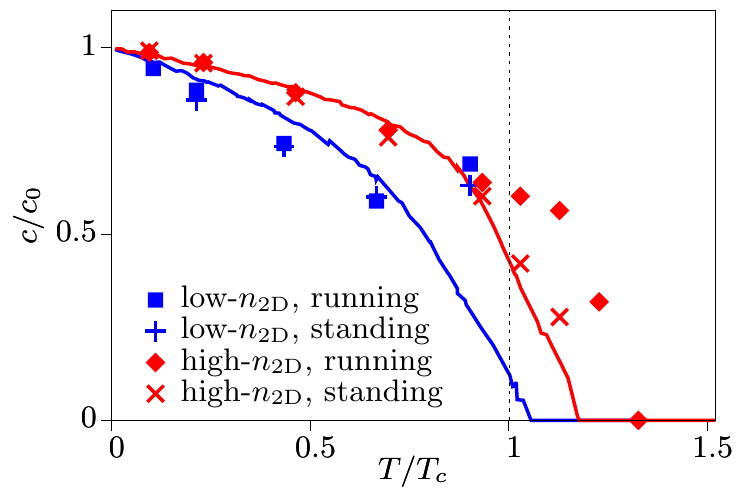}
\caption{\textbf{Comparison of the standing-wave and the running-wave velocity. }
Normalized velocity $c/c_0$ of running and standing wave as a function of $T/T_c$, for low and high $n_\m2D$. 
The estimate $c_\mB(T)$ of Eq. \ref{eq:cB} is shown by the blue and red line for low and high $n_\m2D$, respectively.  }
\label{Fig41}
\end{figure}

\begin{figure*}[]
\includegraphics[width=1.0\linewidth]{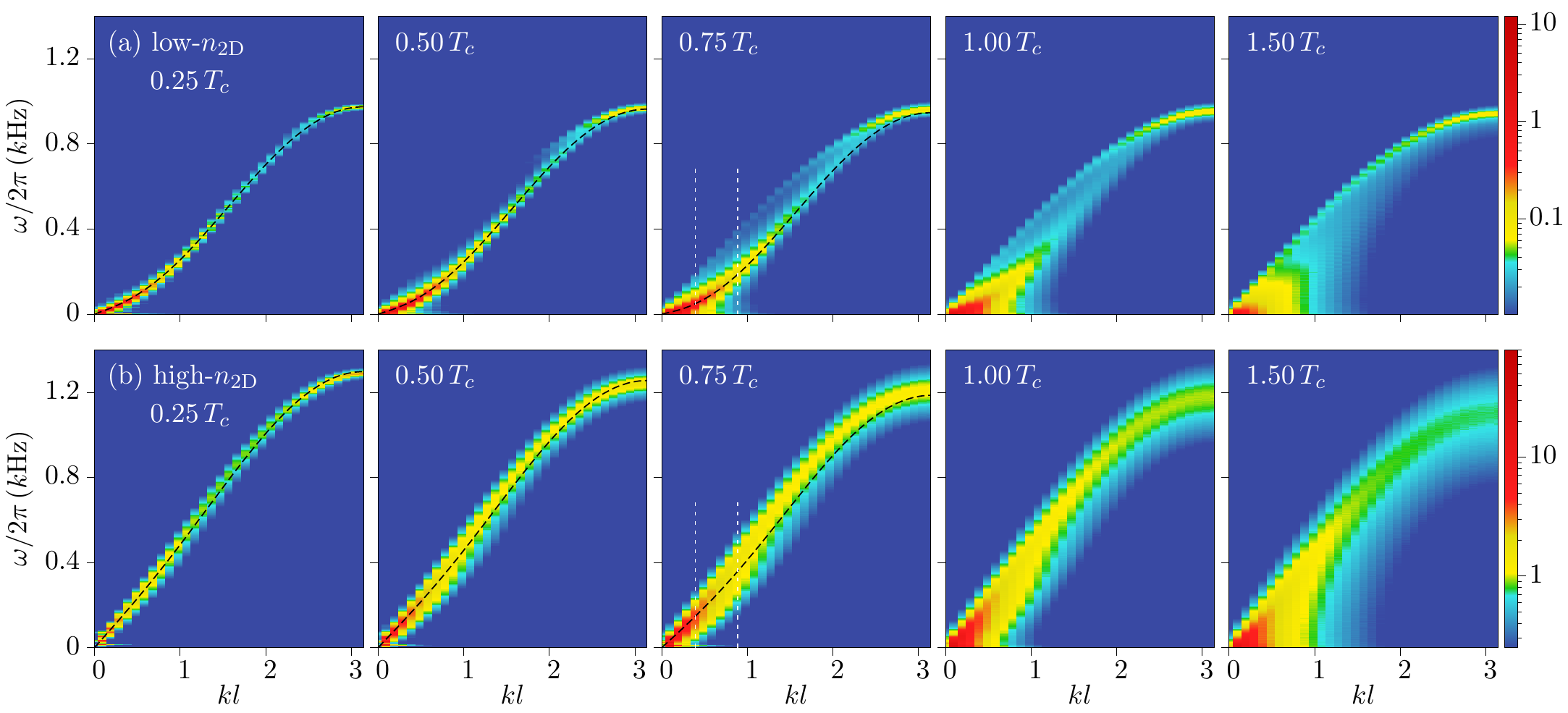}
\caption{\textbf{Excitation spectrum.} 
Dynamic structure factor $S(\bk, \omega)$ as a function of the wavevector $k=k_y$ and frequency $\omega$ is shown throughout the transition, for low $n_\m2D$ (upper row) and high $n_\m2D$ (lower row). 
The black dashed lines are the Bogoliubov dispersion for low and high $n_\m2D$ systems, see text. 
The vertical dotted lines correspond to the fixed-$k$ cuts presented in Fig. \ref{Fig31}. }
\label{Fig3}
\end{figure*}

   As a second measurement, we analyze standing waves for the same system parameters as in Sec. \ref{sec_density}. 
As an illustration, we choose $n_\m2D \approx  27 \, \mum^{-2}$ and $T/T_c =0.23$, 
and create standing waves by periodically modulating the excitation potential, following the scheme described in Sec. \ref{sec_sim}. 
After $1\, \ms$ excitation time, we calculate the squared amplitude $q^2(y,f)$ of the density modulation at varying modulation frequency $f$, see Sec. \ref{sec_sim} for details. 
We show the results of $q^2(y, f)$ determined as a function of $f$ in Fig. \ref{Fig4}(a). 
This response demonstrates the excitation of the first three standing waves at their mode
frequencies.
We fit the spatial dependence of $q^2(y, f)$ with the function $q^2(y, \omega_j) = \sum_{j} B_j^2 \cos^2(k_{j,\omega} y/2)$ to determine the amplitudes $B_j^2$, where $j$ is the mode index.
We show the extracted amplitudes $B_j^2$ of the standing waves in Fig. \ref{Fig4}(b).
We fit $B_j^2$ with a Lorentzian function to determine the mode frequency $f_j$ and the damping rate $\Gamma_j$. 
We show the determined $f_j$ and $\Gamma_j$ in the insets of Fig. \ref{Fig4}(b).
$f_j$ increases linearly with $j$, which demonstrates that the simulated standing waves correspond to the first three lowest-energy spatial modes. 
$\Gamma_j$ also increases linearly with $j$, which is a feature that is consistent with Landau damping.

  We use the lowest-energy standing wave to determine the sound velocity $c= \omega_1 L_y/\pi$ and 
compare it to the running wave measurement in Sec. \ref{sec_density}. 
For the example given in Fig. \ref{Fig4}, we obtain $c=1.48\, \mmms$, which agrees very well with $c = 1.49\, \mmms$ of the running wave 
measurement. We extend the comparison between the two measurements to the low and high $n_\m2D$ systems across the transition.
We use $V_0$ in the range $V_0/\mu = 0.1 - 0.2$ for all simulations.     
We present the results of standing and running wave simulations in Fig. \ref{Fig41}.

  In addition, we compare the simulation results to the Bogoliubov estimate of the sound velocity at nonzero temperature. 
We express $\hat{\psi}$ in the density-phase representation as $\hat{\psi}(\br)= \sqrt{n + \delta \hat{n}(\br)} \exp\bigl( i \hat{\phi}(\br) \bigr) $, where $\delta \hat{n}$ and $\hat{\phi}$ are the density and phase fluctuations, respectively. 
From Eq. \ref{eq_hamil} we obtain the linearized Hamiltonian
\begin{align}\label{eq:Heff}
\hat{\cH}_{0} = \int d \br \, \Big[ \frac{\hbar^2 n_s}{2m} (\nabla \hat{\phi})^2 + \frac{g}{2} (\delta \hat{n})^2 \Big],
\end{align}
where $n_s$ is the superfluid density. The long-wavelength excitations are sound waves with velocity 
\begin{align}\label{eq:cB}
c_\mB = \sqrt{ \frac{g n_s}{m} }.
\end{align}
Following our description in Sec. \ref{sec:term}, this is the second (first) sound estimate for weak (strong) interactions.
We calculate $c_\mB(T)$ by numerically determining $n_s(T)$ using the current-current correlations, see Appendix \ref{app:ns}. 
In Fig. \ref{Fig41} we present the results of $c_\mB(T)$ determined for low- and high-$n_\m2D$ systems. 
$c_\mB(T)$ shows a density-dependent behavior and is nonzero above the transition. 
As mentioned above, the sudden jump of the superfluid density is replaced by a crossover regime due to the finite size of the system. 
Both the running-wave and the standing-wave measurement are consistent with the Bogoliubov estimate below the transition. 
For low densities, both measurements show an upward trend as the temperatures approach the crossover regime. 
As it was demonstrated for the running-wave measurement earlier, this upward trend continues at temperatures above the critical temperature. 
This again suggests the interpretation that for the low density regime the normal sound mode is excited at higher temperatures. For high densities, the measured velocities both show a downward trend above the transition temperature. The standing wave measurement follows the Bogoliubov estimate closely, while the running wave measurement stays at a slightly higher value before it approaches zero as well. 
These measurements indicate that the primary excitation is the diffusive mode, while being slightly sensitive to the specific excitation method.

\section{Dynamic structure factor}\label{sec_dsf} 

\begin{figure}[]
\includegraphics[width=1.0\linewidth]{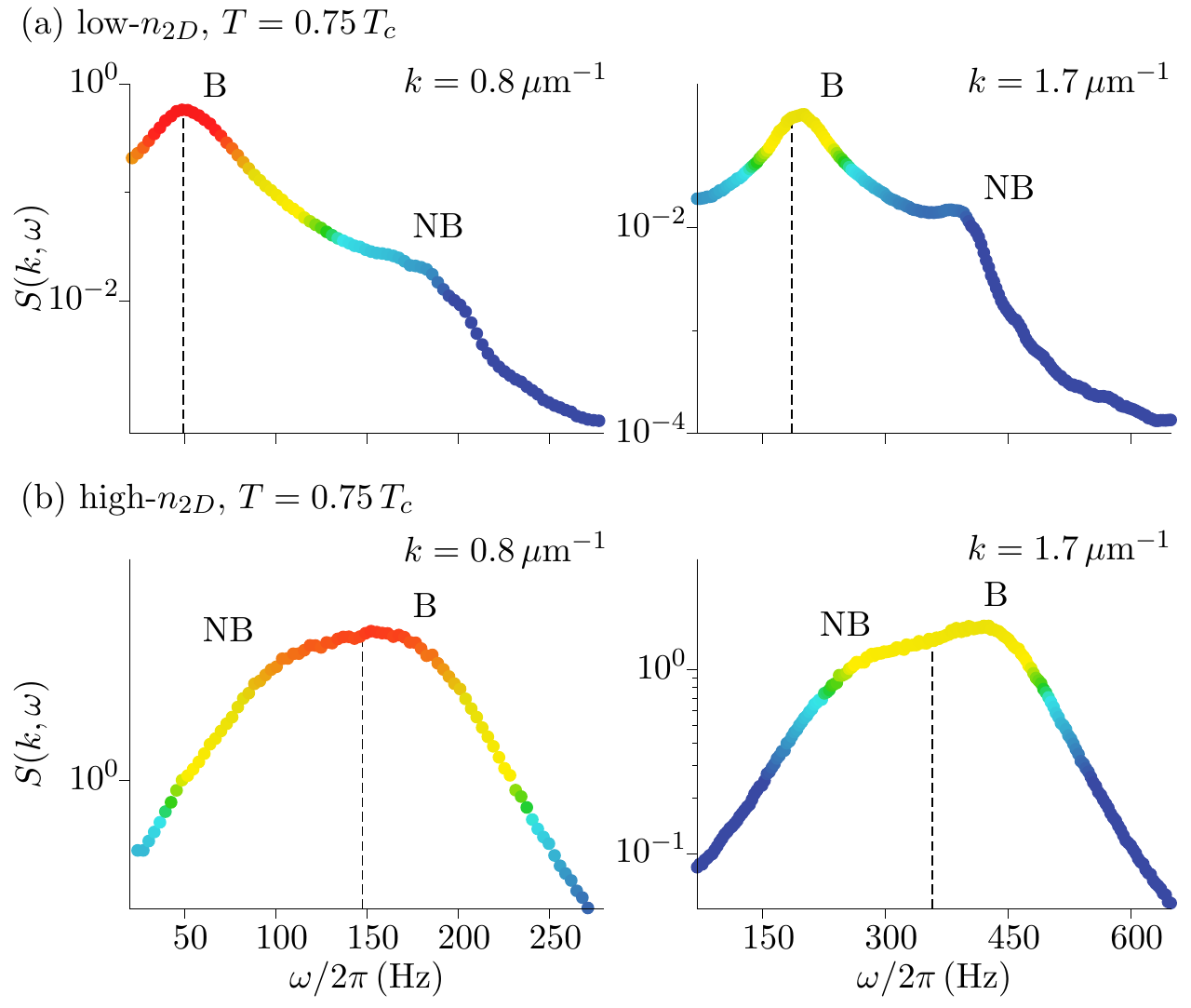}
\caption{$S(\bk, \omega)$ plots at $k = 0.8 \, \mum^{-1}$ and $1.7 \, \mum^{-1}$, for low $n_\m2D$ (upper row) and high $n_\m2D$ (lower row). 
The color scheme is the same as Fig. \ref{Fig3}. 
The vertical dashed lines mark the frequencies of the Bogoliubov dispersion shown in Fig. \ref{Fig3}. }
\label{Fig31}
\end{figure}

\begin{figure*}[]
\includegraphics[width=1.0\linewidth]{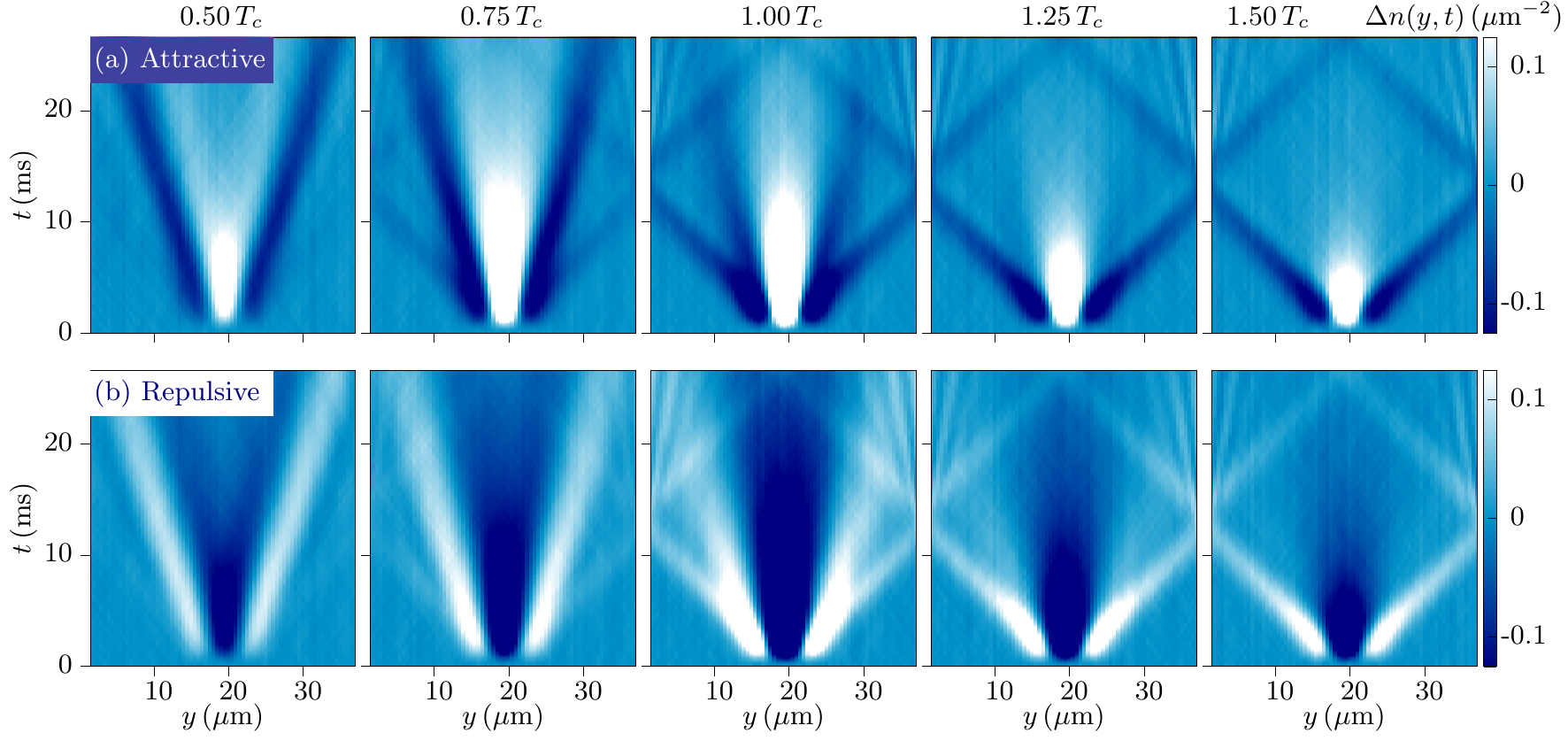}
\caption{\textbf{Excitation of two sound pulses.}
Time evolution of the density profile $\Delta n(y, t)= n(y, t)- n(y, 0)$ shown for (a) attractive and (b) repulsive step-pulse density excitations across the transition. 
We observe the excitation of the Bogoliubov mode at all temperatures below $T_c$,  both modes at intermediate and high temperatures, and the normal sound mode above the transition.  
The Bogoliubov mode transforms into the diffusive mode at $T/T_c=1$. 
The results are obtained for $n_\m2D \approx 3 \, \mum^{-2}$.  }
\label{Fig5}
\end{figure*}

We calculate the dynamic structure factor 
\begin{align}
S(\bk, \omega) =  \langle |n(\bk, \omega)|^2  \rangle,
\end{align}
where $n(\bk, \omega)$ is the Fourier transform of the density $n(\br, t)$ in space and time. We determine $n(\bk, \omega)$ via
\begin{align}
n(\bk, \omega) = \frac{1}{\sqrt{N_l T_s}} \sum_i \int dt \, e^{-i(\bk \br_i - \omega t)} n(\br_i, t).
\end{align}
$N_l$ is the number of lattice sites and $T_s = 328 \, \mms$ is the sampling time for the numerical Fourier transform. 
The dynamic structure factor displays the overlap of the density degree of freedom with the collective excitations. 
We calculate $S(\bk, \omega)$ at various $T/T_c$ for low and high $n_\m2D$. 
In Fig. \ref{Fig3}(a) we show $S(\bk, \omega)$ as a function of the wavevector $k= k_y$ and frequency $\omega$ for low $n_\m2D$ across the transition. 
At low $T$, $S(\bk, \omega)$ has most of its weight at the Bogoliubov branch.
At intermediate $T$, an additional branch with higher velocity appears.
For comparison, we plot the Bogoliubov spectrum $\hbar \omega_{k} = \sqrt{ \epsilon_{k} (\epsilon_{k} + 2m c_\mB^2 ) } $, 
where $c_\mB(T)$ is determined numerically, as above. 
$\epsilon_{k} = 2J \bigl( 1 - \cos(k l) \bigr)$ is the free-particle spectrum on the lattice that is introduced to perform the numerical work, and $J =  \hbar^2 /(2ml^2)$ is the tunneling energy.
This dispersion recovers the continuum dispersion for $l \rightarrow 0$.
The Bogoliubov spectrum agrees well with the lower excitation branch at all $k$, for all $T$ below $T_c$. 
With this, we identify the lower branch as the Bogoliubov (B) mode and the upper branch as the non-Bogoliubov (NB) mode. 
This additional peak is also visible in Fig. \ref{Fig31}(a), where the dynamic structure factor is depicted at two fixed values of the momentum. 
As illustrated in Sec. \ref{sec:term}, the faster mode is the NB mode and the slower mode is the B mode, for this density regime. 
Near $T_c$, the B mode vanishes and becomes the diffusive mode, while the NB mode continuously connects to the normal sound mode of a thermal gas. 
Furthermore, the broadening of the B mode is visible, which corresponds to Landau damping, discussed before.

   In Fig. \ref{Fig3}(b) we show $S(\bk, \omega)$ for a high density $n_\m2D$. 
At low $T$, the weight is again mainly on the Bogoliubov branch, similar to the case of low density. 
At intermediate $T$,  an additional branch with a lower velocity appears, in contrast to the case of low density where the velocity was higher. 
This corresponds to the second scenario described in Sec. \ref{sec:term}. These two branches are also visible in Fig. \ref{Fig31}(b).
We note that the dispersion of the Bogoliubov mode is renormalized to slightly higher values due to level repulsion between the two branches. 
Furthermore, both branches are broadened  more strongly than for low densities, due to the higher interaction. This results in overlapping branches. 
At the transition, the B mode crosses over into the normal sound mode while the second sound mode transforms into the diffusive mode. 
The diffusive mode is broader than for low densities, and has higher weight. This leads to the previous observation that for this regime it is the diffusive mode that is primarily excited with a perturbation of the density.

\section{Excitation of  both sound modes} \label{sec_two_sounds}

\begin{figure}[]
\includegraphics[width=1.0\linewidth]{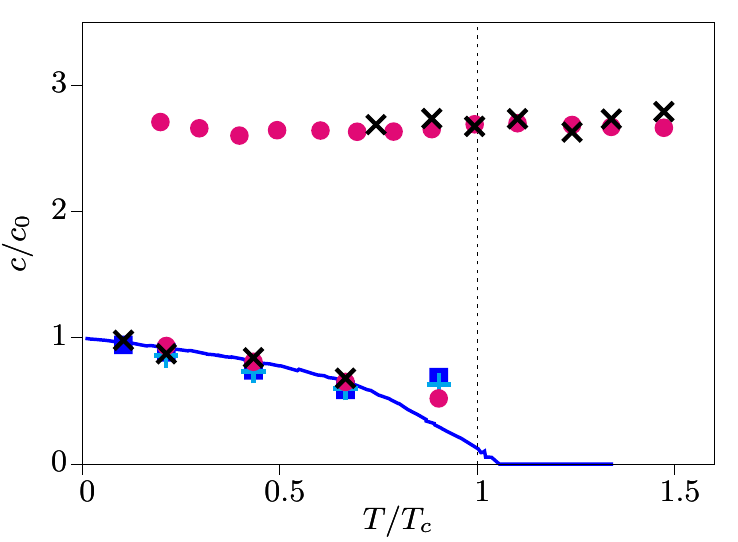}
\caption{\textbf{Sound velocities for low density.} 
Temperature dependence of the two sound velocities of the step-pulse excitation (crosses) and the dynamic structure factor (circles). 
The results of running wave (squares), standing wave (plus symbols), and the Bogoliubov estimate (blue line) are the same as in Fig. \ref{Fig41}.  }
\label{Fig6}
\end{figure}

We propose to excite both sound modes simultaneously by using a step-pulse density perturbation which is created by suddenly turning on and off the Gaussian potential at the location $L_y/2$, see \cite{Denschlag}.
We choose the excitation time to be about $1\, \mms$.
For low-$n_\m2D$ and various $T/T_c$, we excite sound modes using both attractive and repulsive potentials.   
For all simulations, we use $\sigma=2\, \mum$ and $V_0$ in the range $V_0/\mu=0.25-2$.
We show the  results in Fig. \ref{Fig5}.
At $T/T_c=0.5$, the time evolution of the density profile $\Delta n(y, t)$ shows primarily the excitation of the B mode.
We do not observe significant NB mode excitation at and below $T/T_c=0.5$ compared to the numerical noise. 
At higher $T/T_c$, the time evolution shows both B and NB mode excitations which are characterized by two density pulses traveling at different velocities. 
The NB mode travels faster than the B mode. 
At $T/T_c=1$, the B mode transforms into the diffusive mode and the NB mode into the normal sound mode of a thermal gas.
Above $T_c$, the time evolution shows primarily the normal sound propagation, as well as diffusive dynamics at the location of the perturbation. 
We fit the density profile with one or two Gaussians to determine the locations of one or two density pulses.
From these locations we determine the sound velocities.
To cancel out nonlinear effects due to the perturbation potential, we estimate the average squared velocity $c^2= (c^2_\matt + c^2_\mrep)/2$, 
where $c_\matt$ ($c_\mrep$) corresponds to the attractive (repulsive) potential.

     In Fig. \ref{Fig6} we show the temperature dependence of the two mode velocities of the step-pulse excitation.   
For comparison, we determine the mode velocities from the dynamic structure factor of low-$n_\m2D$ shown in Fig. \ref{Fig3}(a).
We fit the excitation spectrum in the low-energy regime with a Lorentzian function to determine the mode frequencies. 
The NB mode frequency is determined after subtracting the background of the B mode.
From the frequencies of the NB and B mode, we determine the first and second sound velocity, respectively.
We show these results for various $T/T_c$ in Fig. \ref{Fig6}. 
The results of the dynamic structure factor show excellent agreement to those of the step-pulse excitation.
Overall, the first sound velocity shows a weak temperature dependence across the transition and is in the range $c/c_0= 2.6 - 2.8$.
The second sound velocity decreases with increasing temperature and vanishes above $T_c$. 
The second sound results are in good agreement with the Bogoliubov estimates and the running- and standing-wave velocities.
This set of results correspond to the first scenario of Sec. \ref{sec:term}.

\section{Conclusions} \label{sec_conc}

  We have studied the propagation of sound in a 2D quasi-condensate of $^{87}$Rb atoms across the superfluid-thermal transition using c-field dynamics.
We have identified two sound modes. To determine one or both velocities of these modes, we employ several methods. The first two methods are inspired by Ref. [33]: we excite running and standing waves with a weak Gaussian potential, from which we obtain a single velocity. Our simulations are in good agreement with the measurements of Ref. [33]. 
Furthermore, we have determined the dynamic structure factor. It displays two sound modes, and provides information about the overlap of these modes with the density degree of freedom. Below the critical temperature, one of the modes is the Bogoliubov mode. We refer to the other mode as the non-Bogoliubov mode. Above the critical temperature, we find the normal sound mode and the diffusive mode. The modes that are detected in Ref. [33] are the Bogoliubov mode below the critical temperature and the normal and the diffusive mode above the critical temperature. 

As a key observation, we find that the non-Bogoliubov mode can have a higher or a lower velocity than the Bogoliubov mode. For weak interactions or low densities, the non-Bogoliubov mode has a higher velocity than the Bogoliubov mode, while for stronger interactions or higher density, the Bogoliubov mode has the higher velocity. While the strongly interacting regime is consistent with a hydrodynamic two-fluid approach, the weakly interacting regime provides a non-hydrodynamic scenario for the collective modes of Bose-Einstein condensates. We propose to measure the two sound modes simultaneously via a step-pulse density perturbation. By choosing the weak and the strong coupling regime of a condensate, these two regimes can be identified, which provides insight into this dynamical regime of condensates.

\section*{acknowledgements}
We thank Jean Dalibard, J\'er\^ome Beugnon, and Sylvain Nascimbene for valuable insights, 
and Ilias M. H. Seifie, Daniel Hoffmann, Johannes Hecker Denschlag, Markus Holzmann and Tilman Enss for stimulating discussions. 
This work was supported by the DFG in the framework of SFB 925 and the excellence clusters `The Hamburg Centre for Ultrafast Imaging’- EXC 1074 - project ID 194651731 and `Advanced Imaging of Matter’ - EXC 2056 - project ID 390715994.

\appendix

\section{Superfluid density}\label{app:ns}

\begin{figure}[b]
\includegraphics[width=1.0\linewidth]{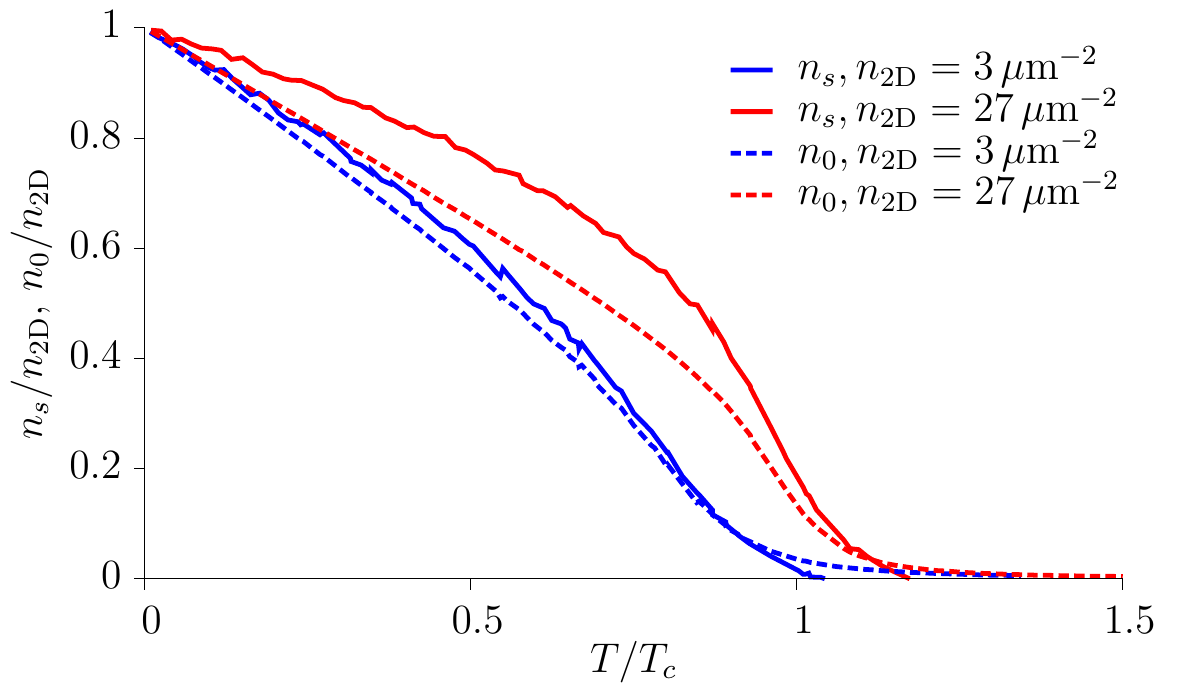}
\caption{\textbf{Superfluid and condensate density.}
Superfluid fraction $n_s/n_\m2D$ and condensate fraction $n_0/n_\m2D$ as a function of $T/T_c$ for low and high-$n_\m2D$.    }
\label{Fig:ns}
\end{figure}

 To determine the superfluid density we calculate the current-current correlations in momentum space. 
The current density $\bj(\br)$ is defined as  
\begin{align}
\bj(\br) = \frac{\hbar}{2im} [  \psi^\ast (\br) \nabla \psi (\br)  - \psi (\br) \nabla \psi^\ast (\br)  ].
\end{align}
By choosing the gradient direction, we calculate the Fourier transform of the current density $(j_\bk)_{x/y}$ in the $x$ and $y$ directions.
We calculate $\langle (j_{\bk}^\ast)_x  (j_{\bk})_y  \rangle $ using equilibrium simulations, with periodic boundary condition.
In the limit $k \rightarrow 0$, these correlations are approximated by (\cite{Stringari, SF_ns})
\begin{align}\label{eq:corr}
\langle (j_{\bk}^\ast)_l  (j_{\bk})_m  \rangle = \frac{k_\mB T}{m} A \Bigl( n_s \frac{k_l k_m}{k^2} + n_n \delta_{lm}  \Bigr).
\end{align}
$n_s$ and $n_n$ are the superfluid and the normal fluid density, respectively. 
$A$ is the system area. 
We analyze the correlations along the line $k_x= k_y = k/ \sqrt{2}$ and determine the $k=0$ value using a linear fit in the low-$k$ regime. 
This allows us to determine $n_s$ at temperature $T$ following Eq. \ref{eq:corr}. 
In Fig. \ref{Fig:ns} we show the determined $n_s$ as a function of $T/T_c$ for low and high $n_\m2D$. 
 $n_s/n_\m2D$ shows a density-dependent behavior and no jump at the transition, due to the finite size of the system. 
 For comparison, we numerically determine the condensate density $n_0$ and show this result in  Fig. \ref{Fig:ns}. 
$n_s$ and $n_0$ show good agreement for low density, while they deviate for high density. 
We note that for finite systems the condensate density scales algebraically with the system size, 
where the scaling exponent is associated with the superfluid density  \cite{Dalibard2011}.

\end{document}